\documentclass[conference,10pt]{IEEEtran}
\makeatletter
\newcommand*{\rom}[1]{\expandafter\@slowromancap\romannumeral #1@}
\makeatother

\IEEEoverridecommandlockouts
\ifCLASSINFOpdf
\else
\fi
\usepackage{lipsum}
\usepackage{graphicx}
\usepackage[utf8]{inputenc}
\usepackage{lmodern,textcomp}
\usepackage{overpic}
\usepackage{graphics}
\usepackage{multicol}
\usepackage{lipsum}
\usepackage{color}
\usepackage[cmex10]{amsmath}
\usepackage{amsthm}
\usepackage{amssymb}
\usepackage{mathrsfs}
\usepackage{mathtools}
\usepackage{amsbsy}
\usepackage{mathrsfs}
\usepackage{setspace}
\usepackage{float}
\usepackage{graphicx}
\usepackage{psfrag}
\usepackage{lettrine}
\usepackage{newclude}
\usepackage[normalem]{ulem}
\usepackage{latexsym}
\usepackage{algpseudocode}
\usepackage{algorithm,algpseudocode}
\usepackage{algorithmicx}
\usepackage{graphicx,amsthm,amsmath,amssymb}
\usepackage{multirow}
\usepackage{caption}
\usepackage{rotating}
\usepackage{booktabs}
\usepackage{amsmath}
\usepackage{wasysym}
\usepackage{mathrsfs}
\usepackage{bbm} % indicator function
\usepackage{filecontents}
\usepackage{amsmath,epsfig,cite,amsfonts,amssymb,psfrag,color}
\usepackage{apptools}
\usepackage{graphicx}
\usepackage[table]{xcolor}
\usepackage{chngcntr}
\usepackage{blindtext}

\usepackage{cite}
\usepackage{amsmath,amssymb,amsfonts}
\usepackage{bm}
\usepackage{algorithm}
\usepackage{algpseudocode}
\usepackage{float}
\usepackage{amsthm}
\usepackage{graphics}
\usepackage{graphicx}
\usepackage{textcomp}
\usepackage{xcolor}
\usepackage{comment}
\usepackage{rotating}
\usepackage{makecell}
\usepackage{multirow}

\theoremstyle{plain}

\newtheorem{proposition}{Proposition}

\def\BibTeX{{\rm B\kern-.05em{\sc i\kern-.025em b}\kern-.08em
    T\kern-.1667em\lower.7ex\hbox{E}\kern-.125emX}}

\usepackage{subcaption}

\begin{document}

\title{Interplay between VAoI, Packet Error Rate, and Delay for Energy-Efficient Remote Monitoring
\thanks{The authors Yasaman Khorsandmanesh and Zinat Behdad contributed equally to this work.}
\thanks{This work was supported by the SweWIN center (2023-00572) funded by Vinnova, Sweden's Innovation Agency.}}

\author{\IEEEauthorblockN{\normalsize
    Yasaman Khorsandmanesh, %\textit{Student Member, IEEE},
        Zinat Behdad, %\textit{Student Member, IEEE},
    Emil Björnson,  %\textit{Fellow, IEEE},
    %\textit{ Member, IEEE}
    and Cicek Cavdar
   }\IEEEauthorblockA{
\textit{Department of Communication Systems, KTH Royal Institute of Technology, Stockholm, Sweden}\\ Email: \{yasamank, zinatb, emilbjo, cavdar\}@kth.se}}

\maketitle

\begin{abstract}
This letter studies energy optimization of short-packet transmission for event-triggered remote monitoring over finite-blocklength wireless links. A wireless sensor node generates updates only when the source state changes, and freshness is measured by the Version Age of Information (VAoI). We model the VAoI evolution as a Markov chain and show its coupling with the packet error rate, characterized by decoding error probability, and average delay. Then, we formulate a transmit-power allocation problem that minimizes the long-term average energy consumption under a VAoI constraint and solve it using a low-complexity search method. Numerical results show that the update arrival probability and blocklength strongly affect the energy--VAoI tradeoff, and that optimizing long-term energy consumption can substantially reduce energy compared with minimizing the energy per transmission.
\end{abstract}

%%%%%%%%%%%%%%%%%%%%%%
\section{Introduction}

Remote monitoring systems are used to track time-varying processes in applications such as environmental sensing, healthcare, traffic monitoring, and industrial automation~\cite{akyildiz2002wireless,rawat2014wireless}. In these systems, wireless sensor nodes (WSNs) send status updates to a monitoring station (MS) to estimate the current source state. Since such updates are often short and time-sensitive, short-packet transmission is a natural transmission framework~\cite{durisi2016toward}. However, in the finite-blocklength regime, the packet error rate, characterized by the decoding error probability (DEP), cannot be neglected. The DEP depends on the transmit power and blocklength~\cite{polyanskiy2010channel}. Therefore, reliability, delay, freshness, and energy consumption must be jointly considered.

Information freshness is commonly measured by the Age of Information (AoI), which quantifies the time elapsed since the generation of the latest received update~\cite{kaul2012realtime,yates2021aoi}. However, in event-triggered remote monitoring, updates are generated only when the source state changes, and the MS is mainly affected by how many source changes have not yet been observed. The Version Age of Information (VAoI), which measures the number of source versions by which the MS lags behind the WSN, is a suitable metric for this setting~\cite{yates2021gossip}. Unlike AoI, VAoI captures the version mismatch between the WSN and MS and does not require detailed knowledge of the source or distortion model.

Energy efficiency is also essential because WSNs are typically battery-powered. Increasing the transmit power improves the decoding success probability and reduces retransmissions, but it also increases the energy consumed per transmission~\cite{anastasi2009energy}. In contrast, reducing transmit power saves energy per transmission but may increase the DEP, thereby increasing both the average delay and VAoI. Hence, minimizing the energy of one transmission is not necessarily equivalent to minimizing the long-term energy consumption of the WSN.

Several previous works have studied AoI-aware power allocation and scheduling to balance freshness and energy consumption in wireless status-update systems~\cite{costa2016age,moltafet2019power}. However, energy-efficient VAoI-aware systems are missing. This letter characterizes the coupling between DEP, average delay, and VAoI in event-triggered remote monitoring with short-packet transmission. Based on this relation, we formulate a VAoI-constrained transmit-power allocation problem that minimizes the long-term average energy consumption of the WSN while accounting for finite-blocklength decoding errors and retransmissions. The problem is solved using bisection search to find the feasible power region and golden-section search to minimize the energy consumption. Numerical results show the impact of the update arrival probability and blocklength, and demonstrate that long-term energy minimization outperforms minimizing the energy of a single transmission.

%%%%%%%%%%%%%%%%%%%%%%%%%
\section{System Model}

\begin{figure}
    \centering
    \includegraphics[width=0.9\linewidth]{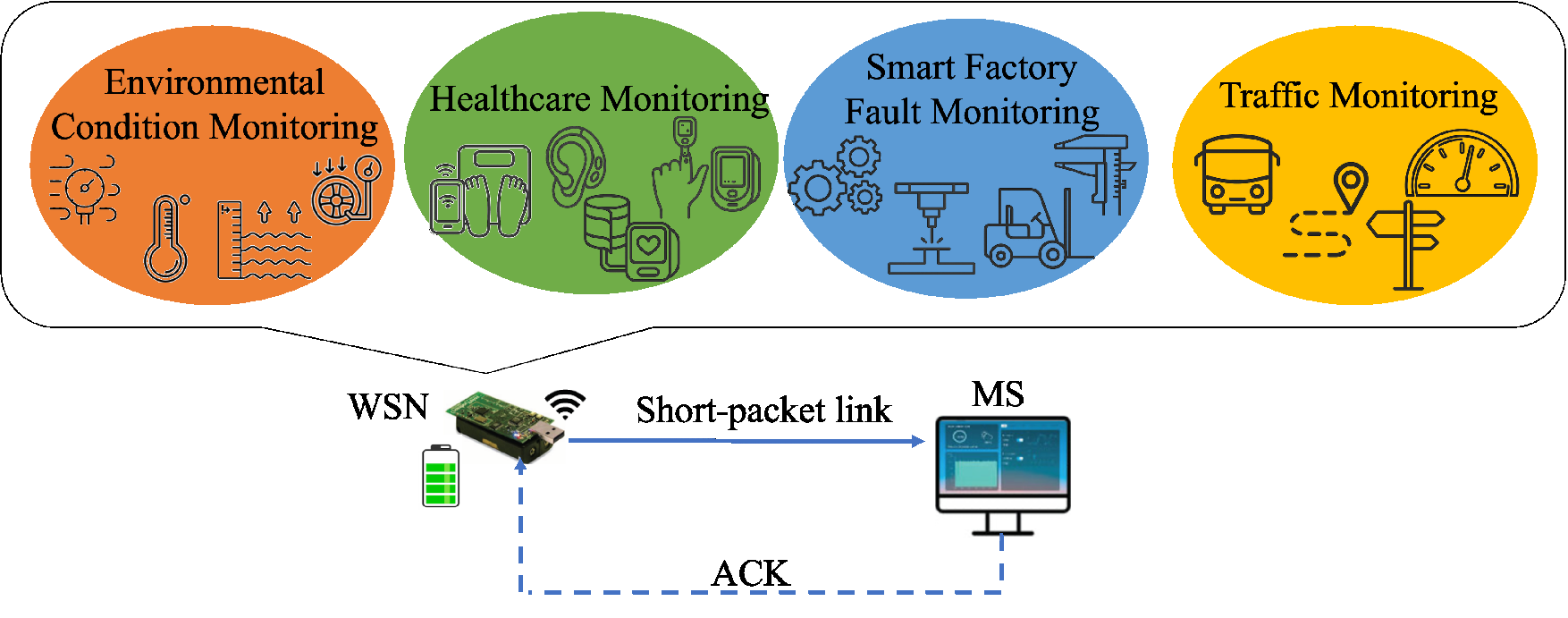}
    \caption{An event-triggered remote monitoring system over a short-packet transmission wireless link.}
    \label{fig:systemmodel}\vspace{-5mm}
\end{figure}

We consider a remote monitoring system in which a WSN monitors a time-varying state and reports status updates to a monitoring station (MS), as illustrated in Fig.~\ref{fig:systemmodel}. The monitored state may represent environmental conditions, patient health indicators, machine fault status, or traffic conditions. Since the updates carry time-sensitive information, they are transmitted over the finite blocklength regime. The MS attempts to decode each received packet and returns an instantaneous ACK via an error-free feedback channel, enabling the WSN to determine whether the update has been successfully delivered.

\vspace{-1mm}
%%%%%%%%%%%%%%%%%%%%%%%%%%%
\vspace{-1mm}
\subsection{Short-Packet Transmission Model}
The considered system targets short-packet transmission for time-critical remote monitoring with short-packet transmission-enabled. In short-packet transmission, the relevant performance metric is the delay until successful reception. Each update contains $b$ information bits and is encoded into a packet of length $L$ channel uses. Since the packets are short, decoding errors cannot be neglected. Let $\gamma_0$ denote the reference channel gain-to-noise ratio, including the effect of path loss, channel gain, and noise power, and $P\in [P_{\min}, P_{\max}]$ denote the WSN transmit power, which is bounded by a minimum and maximum allowable power.
The received signal-to-noise ratio (SNR) at the MS is then modeled as
\begin{equation}
\mathrm{SNR}(P) = P\,\gamma_0.
\end{equation}

For a given transmit power $P$ and finite blocklength $L$, the decoding error probability (DEP) can be approximated in the finite blocklength regime as~\cite{polyanskiy2010channel}
\begin{equation}
\epsilon (P) 
\approx 
Q\!\left(
\sqrt{\frac{1}{V(P) L}}
\left(
L \ln\!\big(1+\mathrm{SNR}(P)\big) 
- b \ln 2
\right)
\right),
\label{eq:epsilon}
\end{equation}
where $Q(\cdot)$ is the Gaussian Q-function and $V(P)=1-\left(1+\mathrm{SNR}(P)\right)^{-2}$ is the channel dispersion. The approximation in~\eqref{eq:epsilon} is accurate for moderate blocklengths and becomes increasingly tight as $L$ grows \cite{polyanskiy2010channel}.

In the considered slotted model, one packet is transmitted over $L$ channel uses. Given the system bandwidth $B$, the slot duration is $T_{\rm s}=\frac{L}{B}$. Hence, the transmission delay of one packet is equal to $T_{\rm s}$. Increasing $L$ improves the decoding reliability in \eqref{eq:epsilon}, but it also increases the duration of each transmission attempt. The delay $D$ is the time elapsed from the first transmission attempt of an update until its successful reception. The average {transmission} delay{for the successfully delivered updates} is 
\begin{equation}
\mathbb{E}\{D\}=\frac{T_{\rm s}}{1-\epsilon(P)}
=\frac{L}{B(1-\epsilon(P))}. \label{eq:delay}
\end{equation}
This expression shows the basic delay-reliability tradeoff: a larger $L$ increases the duration of each attempt, but it can also reduce the DEP and thereby reduce the expected number of attempts.
%%%%%%%%%%%%%%%%%%%%%%%%%%%%%%%%%%%%%%%

\subsection{Status Update Protocol}

The system operates in slotted time, indexed by nonnegative integer numbers $t \in \mathbb{Z}_{\ge0} = \{0,1,2,\ldots\}$. In each slot, the WSN observes the current source state, denoted by $\mathsf{s}_t$. An event-triggered update-generation mechanism is considered, in which a new status update is generated only when the source state changes from the previous slot. The update arrival probability is therefore defined as $\lambda = \Pr\{\mathsf{s}_t \neq \mathsf{s}_{t-1}\}$, which represents the probability that the source state changes between two consecutive slots due to changes in the sensed environment.

The WSN has a buffer of size one. When a state change occurs, the new update is stored in the buffer and becomes the most recent version available at the WSN. If a previous update is still waiting for successful delivery, it is replaced by the newly generated update. This buffer model is natural for remote monitoring, since the MS is mainly interested in the latest state of the source rather than outdated states. If the packet is successfully decoded, the MS updates its stored version of the source and sends an instantaneous ACK signal over an error-free feedback channel. If the transmission fails, the WSN retransmits the latest available update unless it is replaced by a new update in a later slot.
The probability of successful packet decoding in one slot is then
\begin{equation}
\mu(P)=1-\epsilon(P). \label{eq:mu}
\end{equation}
Hence, the transmit power affects the system through the service probability $\mu(P)$. A higher power level provides greater reliability and fewer failed transmissions.

%%%%%%%%%%%%%%%%%%%%%%%%%%%%%%%%
\section{Performance Analysis and Optimization}

In this section, we derive the average VAoI and energy consumption of the considered system. These expressions are then used to formulate the VAoI-constrained energy minimization problem.
%%%%%%%%%%%%%%%%%%%%%%%%%%%%%%%%%%%%%

\subsection{Version Age of Information}

We quantify the timeliness of the status information using the VAoI at the MS. Let $V_{\mathrm{WSN}}(t), V_{\mathrm{MS}}(t) \in \mathbb{Z}_{\ge0}$ denote the source version index available at the WS and the latest version available at the MS in slot $t$, respectively. The VAoI is defined as
\begin{equation}
\Delta_v(t)=V_{\mathrm{WSN}}(t)-V_{\mathrm{MS}}(t).
\end{equation}
Thus, $\Delta_v(t)$ measures how many source state versions the MS is behind the WSN.

\begin{figure}
    \centering
    \includegraphics[width=0.9\linewidth]{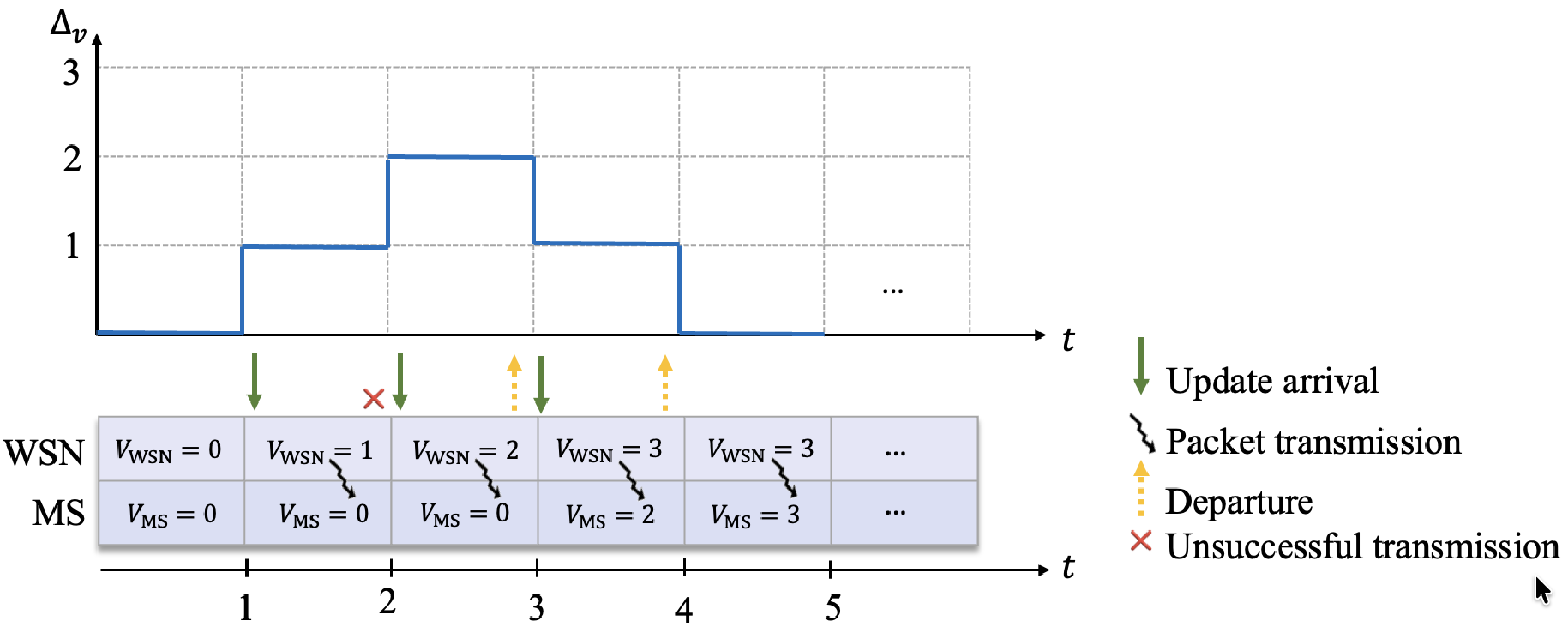}
    \vspace{-2mm}
    \caption{Evolution of VAoI within the system over time.}\vspace{-3mm}
    \label{fig:vaoi}
\end{figure}

The evolution of $\Delta_v(t)$ is governed by the update arrival probability $\lambda$ and the service probability $\mu(P)$ in \eqref{eq:mu}. As illustrated in Fig.~\ref{fig:vaoi}, a new update independently arrives at the beginning of each slot with probability $\lambda$, while a successful transmission occurs during the slot with probability $\mu(P)$. Each source-state change increments the version number at the WSN by one. If the corresponding update is not successfully delivered, the VAoI increases. Conversely, upon successful reception of an update, the MS updates its version number, and the VAoI decreases. When the received update corresponds to the latest version available at the WSN, the VAoI becomes zero.

\begin{figure}[t!]
    \centering
    \includegraphics[width=0.65\linewidth]{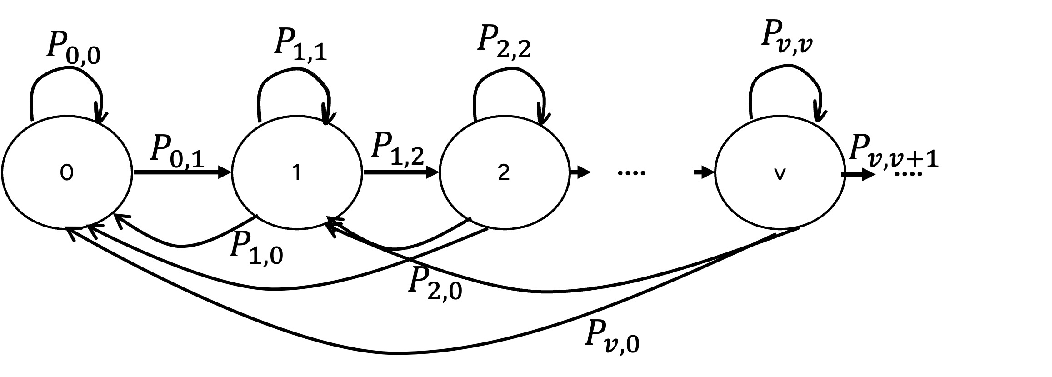}
    \caption{VAoI Markov chain model for the proposed system.}
    \label{fig:markov}\vspace{-5mm}
\end{figure}

The stochastic process $\{\Delta_v(t)\}_{t \in \mathbb{Z}_{\ge0}}$ can be modeled as a discrete-time Markov chain, as shown in Fig.~\ref{fig:markov}, with discrete state space $\mathcal{V}  \in \mathbb{Z}_{\ge0}$, 
where each state represents the version difference between the WSN and MS at the \emph{beginning} of a time slot.

The one-step transition probabilities are determined by the joint effect of arrivals and successful transmissions. From any state $v \geq 0$, the VAoI increases to $v+1$ if a new update arrives at the WSN and the transmission fails. The state remains unchanged if no arrival occurs and the transmission fails. Moreover, the VAoI drops to $0$ or $1$ {from any state $v\geq 1$} depending on whether a successful transmission occurs without or with a new arrival, respectively.

The corresponding transition probability matrix $\mathbf{P}$ is given by \eqref{eq:pmatrix} on the next page.
Let $\Pi_v$ be the steady-state probability that the VAoI Markov chain is in state $v$, i.e.,
\begin{equation}
\Pi_v = \lim_{t\to\infty}\Pr{\Delta_v(t)=v}, \quad v\in\mathcal{V}.
\end{equation}
In particular, $\Pi_0$ is the steady-state probability that the MS has the same version as the WSN, i.e., the MS is not behind the WSN. Hence, the average probability of transmission at the WSN can be defined as the probability of $S >0$, meaning that the $V_{\rm MS}$ is at least one version behind the $V_{\rm WSN}$, and is given by \cite{delfani2026timestamps}
\begin{align}
    1- \text{Pr}\{\mathcal{V}=0\}=1-\Pi_0 =  \frac{\lambda}{\mu(P) + \lambda}.
\end{align}
\begin{figure*}
\begin{equation} \label{eq:pmatrix}
\mathbf{P} = 
\begin{pmatrix}
1-\lambda & \lambda & 0 & 0 & \cdots \\
(1-\lambda)\mu & (1-\lambda)(1-\mu)+\lambda\mu & \lambda(1-\mu) & 0 & \cdots \\
\mu(1-\lambda) & \mu\lambda & (1-\lambda)(1-\mu)+\lambda\mu & \lambda(1-\mu) & \cdots \\
\vdots & \vdots & \vdots & \vdots & \ddots
\end{pmatrix}.
\end{equation}
\hrulefill
\vspace{-4mm}
\end{figure*}

Under the stability condition, $\lambda < \mu(P)$, the Markov chain is positive recurrent and admits a stationary distribution. The resulting average VAoI is
\begin{equation}\label{eq:VAoI}
\Delta_{\rm v} = \frac{\lambda}{\mu(P)},
\end{equation}
where $\mu(P)$ captures the impact of finite blocklength transmission and transmit power on the reliability.

%%%%%%%%%%%%%%%%%%%%%%%%%%%%%%%%

\subsection{Energy Consumption}

We next characterize the energy consumption of the WSN under the considered status update protocol. The energy consumed in a single transmission is
\begin{align}
    E_{\textrm{tr}}=P T_{\rm s},
\end{align}
where $T_{\rm s}$ denotes the duration of one time slot and is a function of the blocklength $L$. Therefore, the transmission energy $E_{\textrm{tr}}$ increases with both $P$ and $L$. 
%\subsection{Total Energy}

However, the long-term energy cost depends not only on the energy cost per transmission but also on the transmission reliability. Increasing either the transmit power or the blocklength improves the successful transmission probability $\mu(P)$, thereby reducing the average number of retransmissions. This reliability gain comes at the cost of higher transmission energy consumption. Therefore, the average long-term energy consumption is determined by the tradeoff between transmission energy and retransmission frequency. Using the steady-state probability $\Pi_0$ that the WSN buffer is empty, the average energy consumption per slot is obtained as
\begin{align}\label{eq:E}
E(P)& = (1-\Pi_0)E_{\rm tr}=\frac{\lambda}{\mu(P) + \lambda} \, P T_{\rm s},
\end{align}
where $\frac{\lambda}{\mu(P) + \lambda}$ represents the fraction of time slots in which the WSN is actively transmitting. 
\subsection{Problem Formulation and Solution}

The objective is to minimize the average energy cost per time slot while guaranteeing a required level of information freshness at the MS. Specifically, we consider the following optimization problem:
\begin{align}
\mathbb{P}_1:\quad\min_{P} \quad & E(P) \\
\text{s.t.} \quad & \Delta_{\rm v}(P) \leq \Delta_{\rm th}, \label{eq:versionthresh}\\
& P_{\min} \leq P \leq P_{\max}.
\end{align}
$E(P)$ is generally unimodal over the feasible interval due to the opposing effects of power increase: higher $P$ increases the instantaneous energy expenditure but reduces the retransmissions through improved reliability. The VAoI constraint in \eqref{eq:versionthresh} imposes upper bounds on both the average delay and the DEP, as stated in Proposition~1, since these quantities are coupled through $\mu(P)$.

\begin{figure}
    \centering
\includegraphics[width=0.55\linewidth]{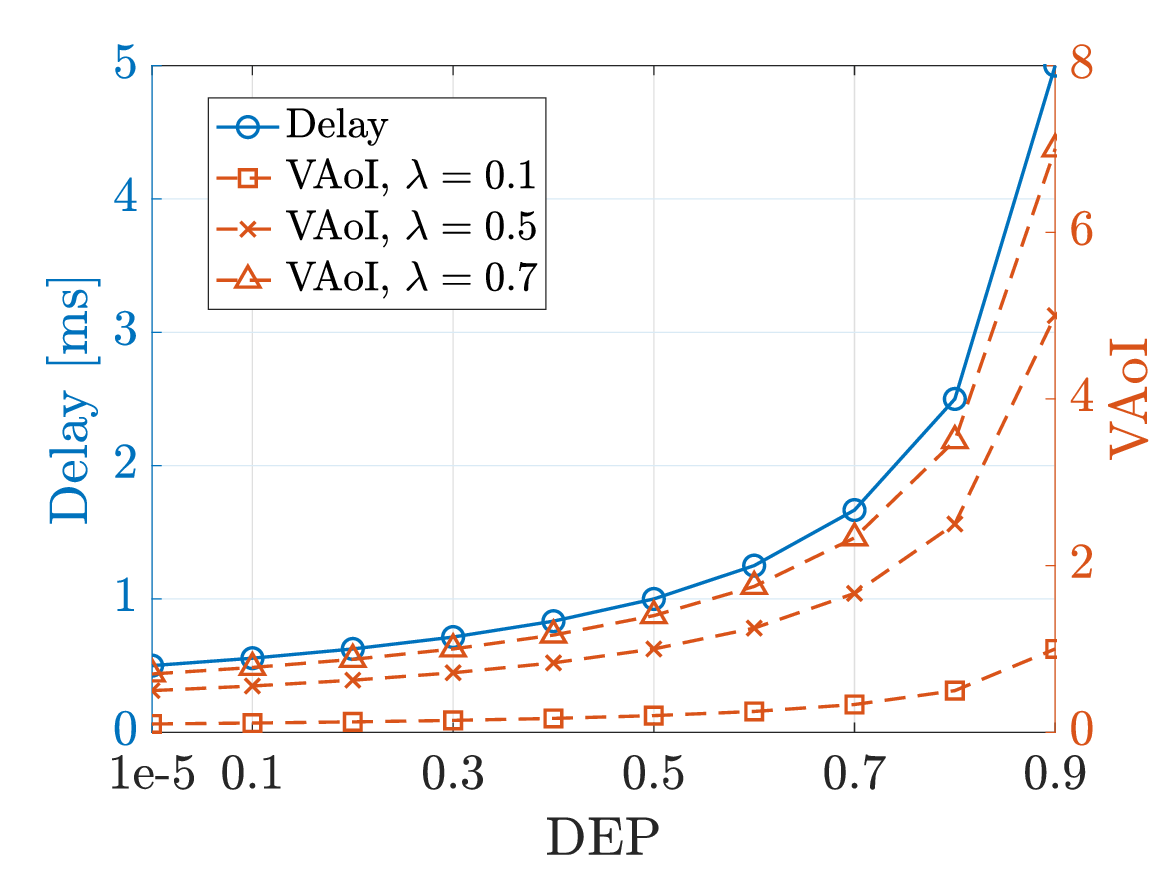} 
   \caption{Delay and VAoI vs. DEP $\epsilon$.}\vspace{-5mm}
    \label{fig:Delay_VAoI}
\end{figure}

\vspace{-2mm}
\begin{proposition}
A higher DEP increases VAoI by reducing the success probability, thereby increasing both retransmissions and the number of source versions by which the MS lags behind the WSN. Following  \eqref{eq:versionthresh}, \eqref{eq:delay} and \eqref{eq:VAoI}, the upper bound on the average delay {of the successfully delivered updates} would be \begin{equation} \mathbb{E}\{D(P)\} \leq D_{\max} \overset{\Delta}{=} \frac{T_{\rm s}\Delta_{\rm th}}{\lambda}. \label{eq:delay_bound_vaoi} \end{equation} From \eqref{eq:VAoI}, the VAoI constraint requires the DEP to satisfy \begin{equation} \epsilon(P) \leq \epsilon_{\max} \overset{\Delta}{=}1-\frac{\lambda}{\Delta_{\rm th}}, \label{eq:epsilon_bound_vaoi} \end{equation} provided that $\Delta_{\rm th}\geq \lambda$. 
\end{proposition}
Fig.~\ref{fig:Delay_VAoI} illustrates the coupling between the DEP, the average delay, and the VAoI, as characterized in Proposition~1. As the DEP $\epsilon$ increases, $\mu(P)$ decreases. Hence, more retransmission attempts are needed before an update is successfully delivered, increasing the average delay. Reliability degradation also increases VAoI. Therefore, delay and VAoI follow the same increasing trend with respect to $\epsilon$. Moreover, higher update arrival probabilities $\lambda$ lead to larger VAoI, since the MS falls behind the WSN by more source versions when state changes occur more frequently. This confirms that the VAoI constraint indirectly limits both the maximum tolerable delay and the maximum tolerable DEP, as shown in Proposition~1.

Problem $\mathbb{P}_1$ is a one-dimensional constrained optimization problem in $P$. 
We solve it in two steps: First, we determine the minimum feasible transmit power that satisfies the VAoI constraint, i.e., the smallest $P$ such that $\Delta_{\rm v}(P) \leq \varepsilon_{\max}$. This can be obtained via bisection search due to the monotonicity of $\Delta_{\rm v}(P)$.  Second, over the feasible interval $[P_{\rm f}, P_{\max}]$, we minimize $E(P)$ using the golden-section search algorithm, which is well suited for unimodal functions without requiring derivative information \cite{pronzato1998generalized}. The detailed procedure is summarized in Algorithm~1.  This approach provides an efficient, low-complexity method to balance energy efficiency and information freshness at the WSN while explicitly accounting for the reliability of finite-blocklength transmission.

\begin{algorithm}[t]
\caption{VAoI-Constrained Energy Minimization}\label{algo:1}
\begin{algorithmic}[1]
\State \textbf{Input}: $[P_{\min},P_{\max}]$, $\delta$, $\lambda$, $\Delta_{\rm th}$, $L$, $\gamma_0$, $b$
\State \textbf{Define}: $\Delta_{\rm v}(P)=\lambda/\mu(P)$ and $E(P)$ from \eqref{eq:E}

\Statex \textit{Step 1: Feasible power interval}
\State $P_{\rm low}\gets P_{\min}$, $P_{\rm high}\gets P_{\max}$
\While{$P_{\rm high}-P_{\rm low}>\delta$}
\State $P_{\rm mid}\gets (P_{\rm low}+P_{\rm high})/2$
\If{$\Delta_{\rm v}(P_{\rm mid})>\Delta_{\rm th}$}
\State $P_{\rm low}\gets P_{\rm mid}$
\Else
\State $P_{\rm high}\gets P_{\rm mid}$
\EndIf
\EndWhile
\State $P_{\rm f}\gets P_{\rm high}$

\Statex \textit{Step 2: Energy minimization}
\State $\phi\gets(\sqrt{5}-1)/2$, $a\gets P_{\rm f}$, $c\gets P_{\max}$
\While{$c-a>\delta$}
\State $P_1\gets c-\phi(c-a)$, $P_2\gets a+\phi(c-a)$
\If{$E(P_1)<E(P_2)$}
\State $c\gets P_2$
\Else
\State $a\gets P_1$
\EndIf
\EndWhile

\State $P^\star\gets(a+c)/2$
\State \Return $P^\star$, $E(P^\star)$, $\Delta_{\rm v}(P^\star)$
\end{algorithmic}
\end{algorithm}

%%%%%%%%%%%%%%%%%

\section{Numerical Results}

\begin{figure}[t]
    \begin{subfigure}{1\linewidth}
        \centering
    \includegraphics[width=0.55\linewidth]{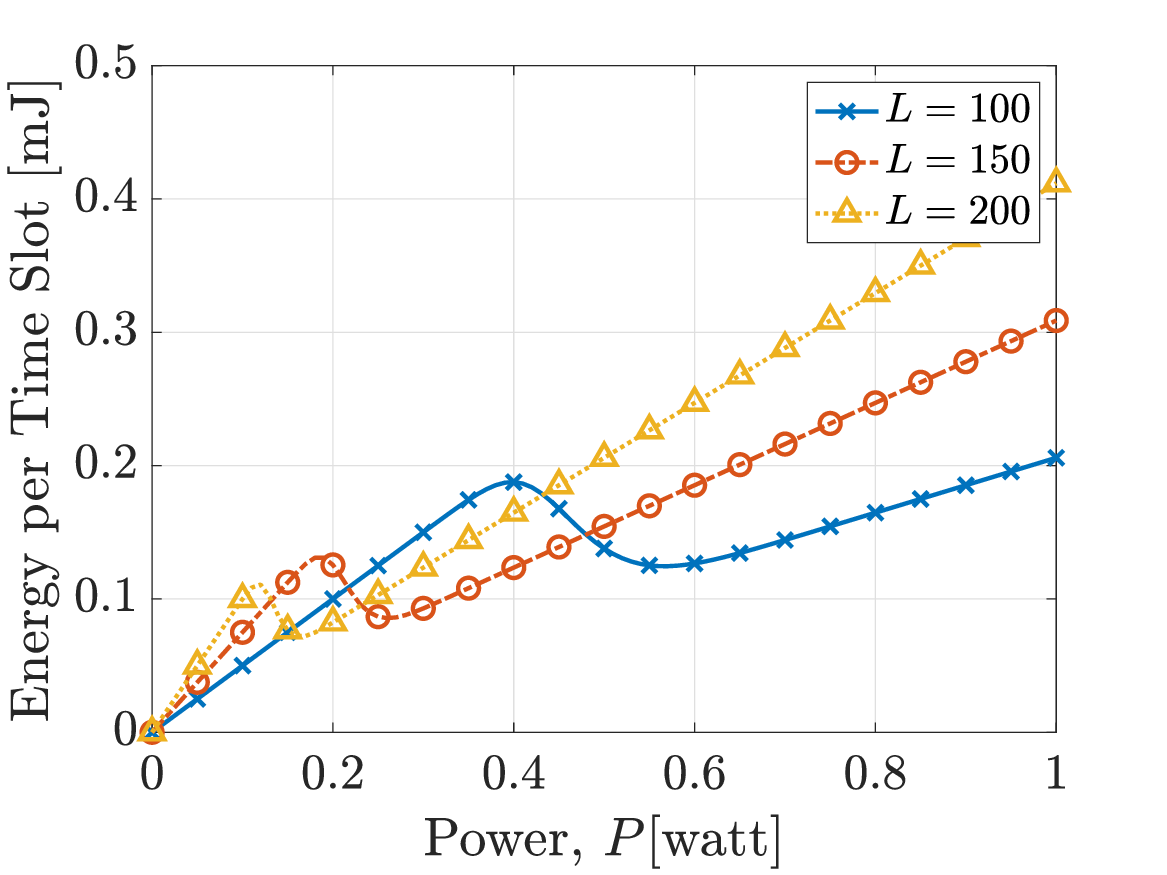} 
   \caption{} \label{fig:1a}
        \end{subfigure}\vspace{-0.5mm}
        \hfill
          \begin{subfigure}{1\linewidth}
    \centering
   \includegraphics[width=0.55\linewidth]{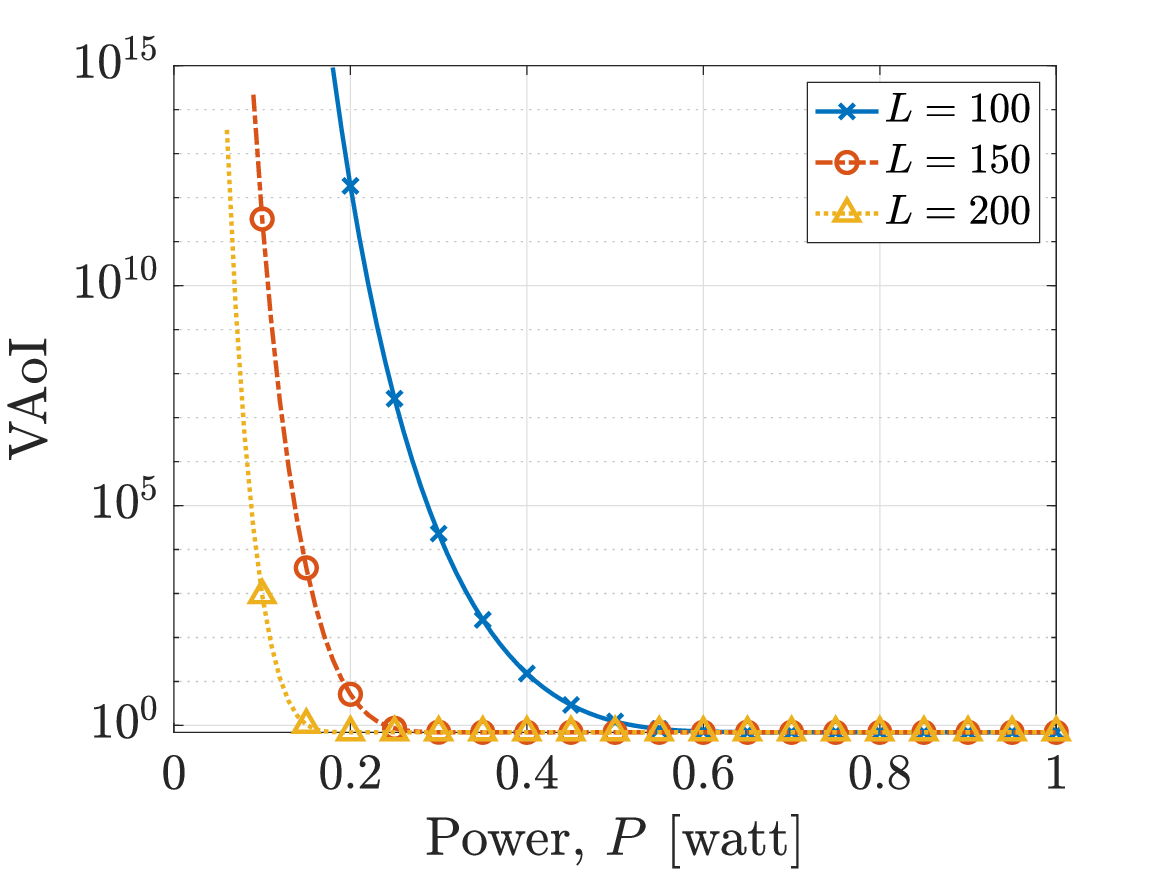} 
   \caption{} \label{fig:1b}
   \end{subfigure}
 \caption{(a) Average energy consumption per time slot, and (b) VAoI vs. transmit power $P$ with $\lambda=0.7$.} \label{fig:1}\vspace{-5mm}
    \end{figure}

In this section, we evaluate the performance of the proposed power-control scheme. Unless otherwise stated, we set $\gamma_0\!=\!10$~W$^{-1}$, coherence bandwidth $B_{\rm c}\!=\!20$ MHz, packet size $b=256$ bits, and $\lambda=0.7$. {The minimum and maximum transmit power are $P_{\max}\!=\!1$~W and $P_{\max}\!=\!1$~W, respectively} and $\delta=10^{-5}$ in Algorithm~\ref{algo:1}. The blocklength $L$ is selected to satisfy the short-packet transmission latency budget through $T_{\rm s}\!=\!L/B$.

Fig.~\ref{fig:1} shows the impact of transmit power on the average energy consumption per time slot and the VAoI for blocklengths of 100, 150, and 200 channel uses. As shown in Fig.~\ref{fig:1a}, for a short blocklength, e.g., $L=100$, low transmit power results in poor reliability and consequently severe freshness degradation. As the transmit power increases, the decoding reliability improves, leading to a substantial reduction in VAoI, as shown in Fig.~\ref{fig:1b}. In contrast, for larger blocklengths, i.e., $L=150$ and $L=200$, the VAoI remains low even at moderate power levels because the longer packets provide higher decoding reliability. Consequently, these cases achieve lower average energy consumption per time slot. Increasing the transmit power increases the energy consumption per transmission but also improves the decoding probability, thereby reducing the number of retransmissions and resulting in lower average energy consumption per time slot. However, when the transmit power exceeds 0.5\,W, energy consumption increases almost linearly with negligible gain in VAoI.

\begin{figure}
    \centering
   \includegraphics[width=0.55\linewidth]{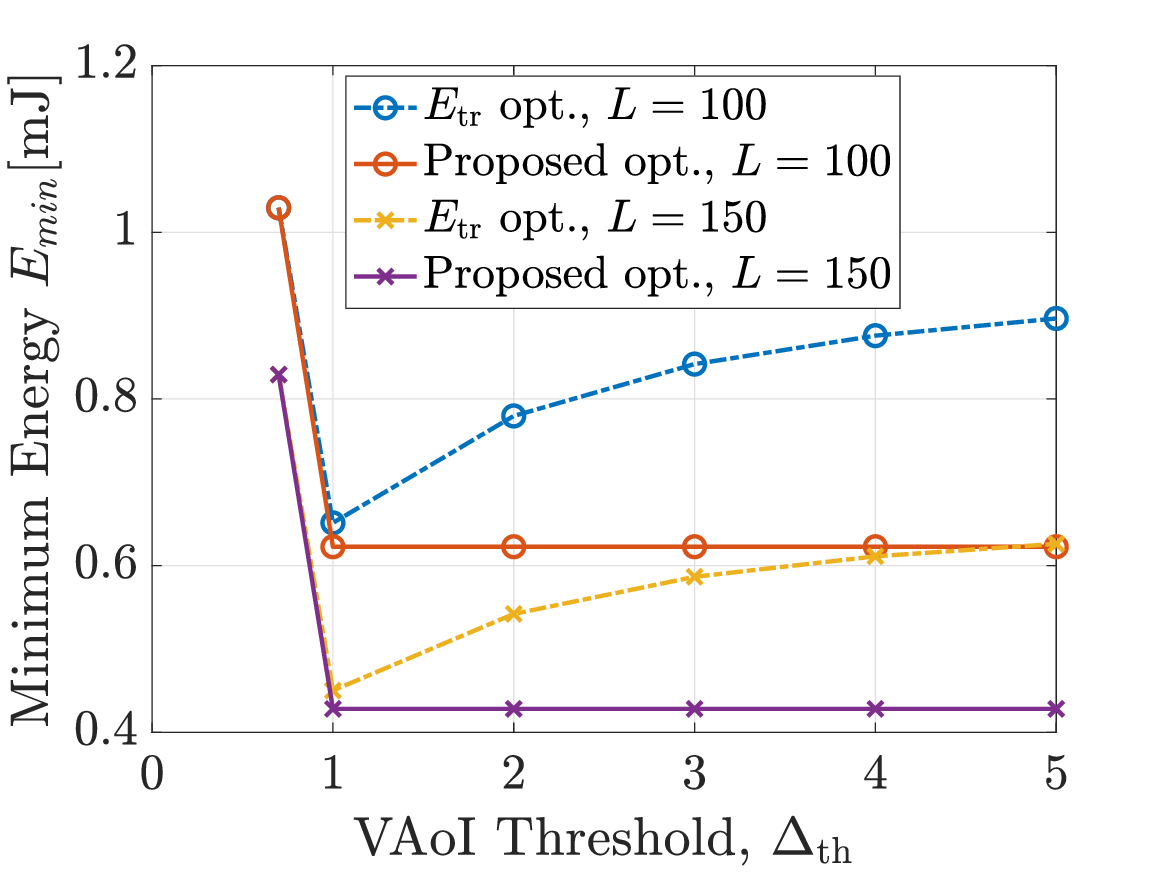} 
   \caption{Minimum energy vs. $\Delta_\textrm{th}$ with $\lambda=0.7$.} 
    \label{fig:4}\vspace{-5mm}
\end{figure}
Fig.~\ref{fig:4} shows the minimum average energy consumption per time slot versus $\Delta_{\rm th}$. We compare the \emph{Proposed opt.} approach with a benchmark, denoted by \emph{$E_{\rm tr}$ opt.}, that minimizes the energy consumption per transmission. When the freshness requirement is strict, i.e., small $\Delta_{\rm th}$, higher transmit power is needed to satisfy the VAoI constraint, resulting in increased energy consumption. Note that the minimum achievable average VAoI is equal to $\lambda$ and is attained when the transmission success probability is one $\mu=1$. As $\Delta_{\rm th}$ increases, the required power decreases and the minimum energy is reduced. Beyond a certain threshold, however, the energy consumption of the proposed scheme saturates, indicating that the optimum is primarily governed by the energy–reliability trade-off rather than the VAoI constraint. In contrast, the benchmark scheme becomes increasingly inefficient at larger $\Delta_{\rm th}$, as it reduces the transmit power without accounting for the resulting increase in retransmissions. Consequently, it fails to fully exploit the additional flexibility provided by a relaxed freshness requirement.

\begin{figure}[t]
    \centering
   \includegraphics[width=0.55\linewidth]{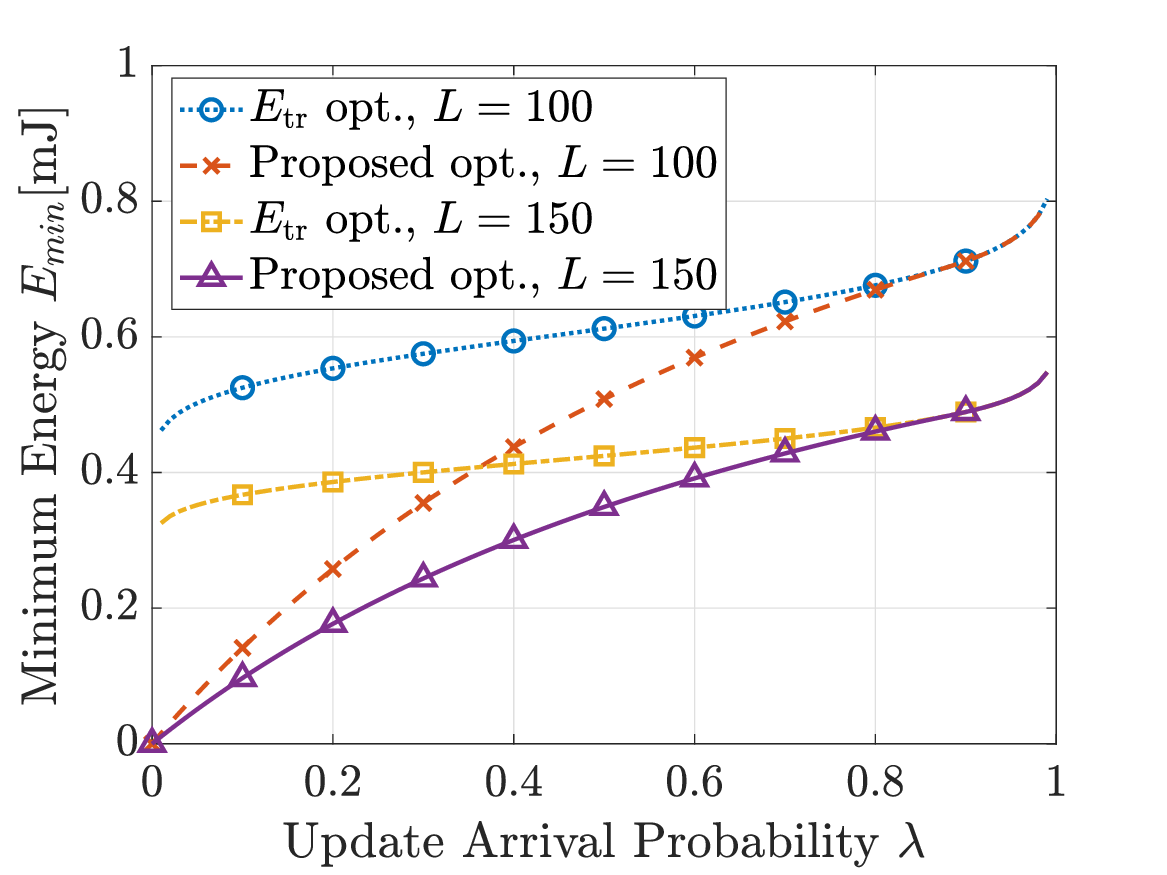} 
   \caption{Minimum energy vs. $\lambda$ with $\Delta_\textrm{th}=1$.}
    \label{fig:3}\vspace{-4mm}
\end{figure}

Fig.~\ref{fig:3} shows the minimum energy consumption as a function of $\lambda$. For all cases, the minimum energy consumption increases with $\lambda$, since more frequent state changes require more updates to satisfy the VAoI constraint. The proposed scheme outperforms the benchmark till $\lambda < 0.8$, with energy savings of up to $96\%$ at $\lambda=0.01$ and $73\%$ at $\lambda=0.1$. For higher $\lambda$, both methods converge as frequent updates tighten the freshness requirement, leaving less flexibility for energy optimization. 
Therefore, minimizing only transmission energy is generally suboptimal, since it neglects the effect of reliability and retransmissions. The gain is particularly pronounced for short blocklengths, where the power--reliability tradeoff is more stringent. For large $\lambda$, however, the \emph{\(E_{\rm tr}\) opt.} scheme with $L=150$ can consume less energy than the \emph{Proposed opt.} scheme with $L=100$, highlighting the impact of blocklength selection.

%%%%%%%%%%%%%%%%%%%

\section{Conclusion}

We studied VAoI-constrained power allocation for event-triggered remote monitoring over finite-blocklength wireless links. By modeling the VAoI evolution as a function of the update arrival probability and the decoding success probability, we formulated a long-term energy minimization problem under an average freshness constraint. The proposed power-control method balances the increased energy per transmission at high power against the retransmission cost caused by decoding failures at low power. It is also demonstrated that DEP and delay can be incorporated into the VAoI framework. The results show that the proposed scheme reduces the average energy consumption compared to transmission-energy minimization, while satisfying the required VAoI constraint. However, when $\lambda$ is large, transmission-energy minimization with a longer blocklength can consume less energy than the proposed scheme with a lower blocklength.

%%%%%%%%%%%%%%%%%%%%%%%%%%
\bibliographystyle{IEEEtran}
\bibliography{IEEEabrv.bib,refs.bib}
\end{document}